\documentclass[aps,groupedaddress,showpacs,letterpaper,twocolumn]{revtex4}
\usepackage{graphicx} 
\usepackage{amsmath}
\usepackage{color}  

\begin{document}

\newcommand {\edit}[1]{\textcolor{red}{#1}}

\title{Efficient Grover search with Rydberg blockade}

\author{Klaus M{\o}lmer$^1$, Larry Isenhower$^2$ and Mark Saffman$^2$}
\affiliation{$^1$Lundbeck Foundation Theoretical Center for Quantum System Research, Department of Physics and Astronomy,
University of Aarhus, DK-8000 \AA{}rhus C, Denmark\\
$^2$Department of Physics,
University of Wisconsin, 1150 University Avenue, Madison, Wisconsin 53706, USA.}
 \date{\today}

\begin{abstract}
We present efficient  methods to implement the quantum computing Grover search algorithm using the Rydberg blockade interaction.
We show that simple $\pi$-pulse excitation sequences between ground and Rydberg excited states readily produce the key conditional phase shift and inversion-about-the-mean unitary operations for the Grover search. Multi-qubit implementation schemes suitable for different properties of the atomic interactions are identified and the error scaling of the protocols with system size is found to be promising for experimental investigation.
\end{abstract}

\maketitle

\section{Introduction}

The last few years have witnessed impressive progress in attempts to perform quantum computing with trapped neutral atoms  via the strong, long range Rydberg blockade interactions \cite{Gaetan2009,Urban2009,IsenhowerPRL2010,Wilk2010,RMP2010}, and research has begun to scale the physical systems to quantum registers encoded in several trapped atoms and to perform non-trivial algorithms. The Rydberg blockade gate was proposed in \cite{Jaksch2000} as a robust and fast quantum gate, relying on the presence of a single atom in a high lying Rydberg state shifting the Rydberg excited state energy of any other nearby atom and thus preventing the resonant excitation of that atom. As the first (control) atom can be selectively excited from any of the ground states representing qubit states $|0\rangle$ and $|1\rangle$, the evolution of the second (target) atom is effectively controlled by the quantum state of the first one.

With the capability to perform suitable one- and two-bit gates, one can decompose any quantum computation as a suitable sequence of such gates. The Rydberg blockade interaction mechanism has a special property that we will make use of in the present communication: A single Rydberg excited atom can simultaneously block the excitation of a whole ensemble of atoms in its vicinity. This collective blockade has been suggested as a means to use an ensemble of atoms to code a single qubit in \cite{Lukin2001}, to generate a wider class of entangled multi-atom states \cite{Moller2008,Saffman2009}, and to code multi-bit quantum registers in ensembles of multi-level atoms \cite{Brion2007,Saffman2008}. In this article we shall show that also in the more conventional encoding with one qubit per atom, the simultaneous blockade among many atoms offers interesting prospects for unique multi-bit quantum gates. In particular, we propose means to carry out the Grover search algorithm \cite{Grover97} on a $k$-bit register, encoded in $k$ individually trapped atoms.

The outline of the paper is as follows. In Sec. II, we discuss the theoretical Grover search algorithm, we describe the Rydberg blockade gate mechanism, and we show formally how the two key operations in the Grover algorithm can both be implemented efficiently by sequences of $\pi$-pulse excitations between the qubit levels and the Rydberg states. In Sec. III, we discuss possible alterations of the gate sequences suited to the case of atoms with different interaction properties. In Sec. IV, we estimate the expected fidelity of the gate operations proposed in the manuscript. Sec. V concludes the paper.

\section{Grover algorithm}

The Grover search algorithm is able to identify with high probability a single marked element $x_0$ out of $N$ candidates with only $\sqrt{N}$ queries of a database or a suitable oracle device. The algorithm assumes that a register, holding a superposition of states $\sum_x c_x |x\rangle$ corresponding to the different possible values $x$ of the unknown variable, is queried by being acted upon with a unitary operation that leaves all components unchanged, except $|x_0\rangle$ which undergoes a change of sign, $c_{x_0} \rightarrow -c_{x_0}$. The query process thus yields a conditional phase shift and does not change the weight of the desired state component within the superposition, and hence it does not change its probability to reveal itself in a measurement on the register.

The key ingredient in the Grover algorithm \cite{Grover97} is the subsequent step, which replaces all amplitudes $c_x$ by their values reflected in their mean value $\overline{c}=\frac{1}{N}\sum_x c_x$:
\begin{equation}
\label{GroverInv}
c_x \rightarrow \frac{1}{N} \sum_{x'} c_{x'} - (c_x-c_{x'}).
\end{equation}
It is clear that if all amplitudes are initially identical with the value $1/\sqrt{N}$, after the conditional change of sign and the reflection in the mean value, the magnitude of the marked element has changed. For a large value of $N$, one gets $c_{x_0} \sim 3/\sqrt{N}$ while all other amplitudes are still of the order $1/\sqrt{N}$ after the first step of the algorithm, and repeated action of the steps will further increase the amplitudes of the marked element by of the order $1/\sqrt{N}$ in each iteration. The $|x_0\rangle$ amplitude approaches unity after $\sim \sqrt{N}$ steps \cite{Grover97}.

\subsection{Query of the oracle with the Rydberg blockade interactions}

\begin{figure}
 \includegraphics[width=8cm]{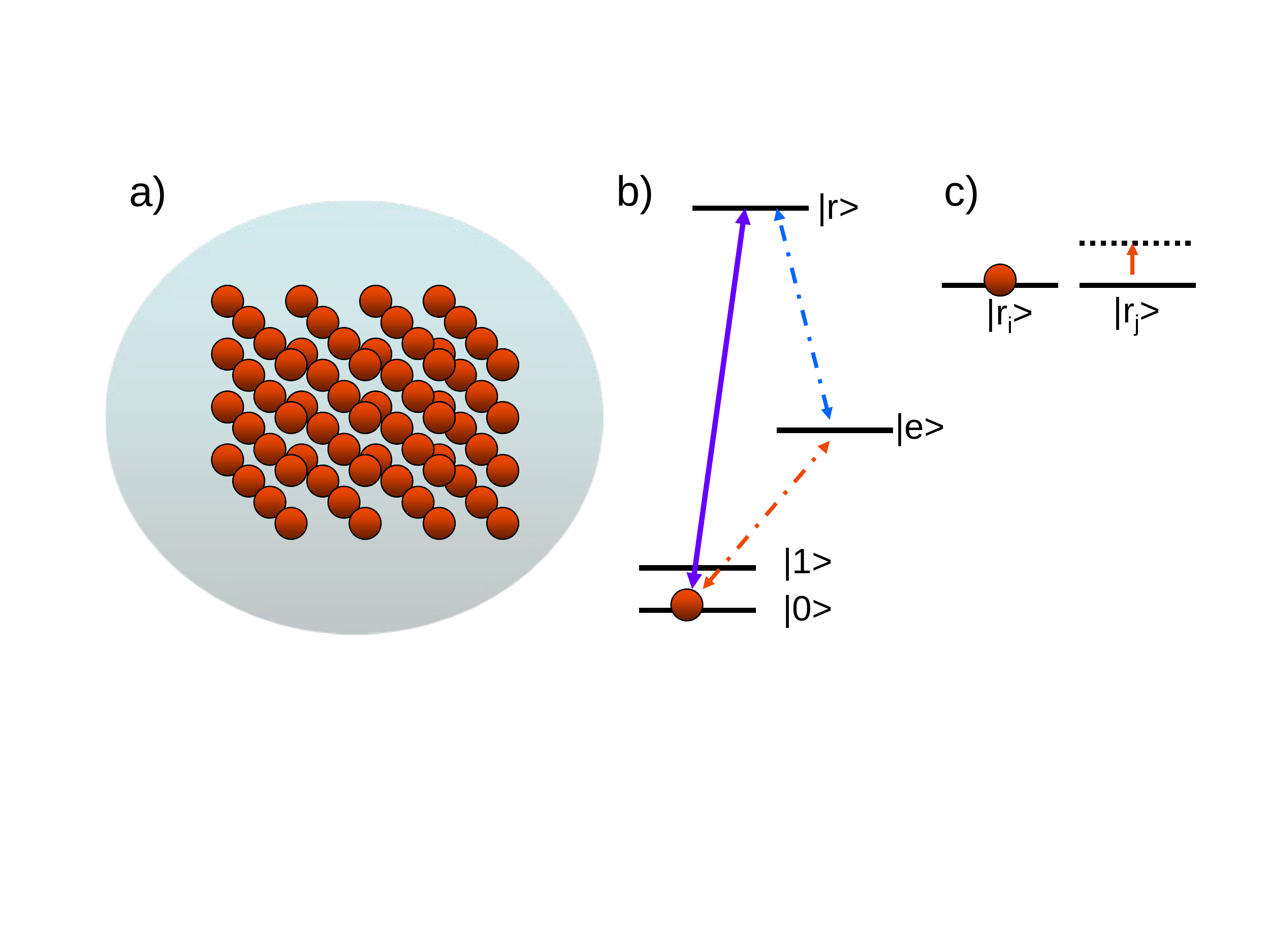}\\
  \caption{Atomic configuration for quantum computing with Rydberg blockade gates. a) An ensemble of atoms individually trapped with mutual distances smaller than the Rydberg interaction blockade radius. b) Level scheme of individual atoms, showing two stable states $|0\rangle$ and £$|1\rangle$ used for qubit storage, and a Rydberg excited state $|r\rangle$ which may be excited either directly or, more conveniently (dashed arrows), by a two-photon transition via the optically excited state $|e\rangle$. c) The blockade interaction is due to the the dipole-dipole interaction which shifts the Rydberg excitation energy of atom $j$ if atom $i$ is already excited.}\label{Fig.1}
\end{figure}

We now assume that the register is represented by $k=\log_2N$ individually trapped atoms, see Fig1.a), with two low-lying long lived states $|0\rangle$ and $|1\rangle$, and a Rydberg excited state $|r\rangle$, see Fig.1.b). The arguments $x$ of the database search are given in a binary representation, and in particular the marked element $x_0$ has the binary representation $b_0,b_1, ... b_{k-1}$, with $b_i=0$ or $1$, represented by atom $i$ occupying state $|0\rangle$ and $|1\rangle$, respectively. To certify whether a candidate $x$ equals $x_0$ one must therefore check that every bit value in the representation of $x$ coincides with the corresponding $b_i$ in $x_0$. In an experimental implementation, we imagine that we hand over the physical system to the oracle -  another component of the experiment, delegated to this role. The oracle must then perform the unitary operation, which puts a change of sign on the $x_0$ component and leaves all other state components unchanged.

Here is how this can be done: the oracle component of the experiment is in possession of the values $b_0,b_1, ... b_{k-1}$ and can thus apply operations on the individual atoms, where the ground state amplitude in $|1-b_i\rangle = |0_i\rangle$ or $|1_i\rangle$ is transferred by a resonant $\pi$-pulse to the Rydberg excited state $|r_i\rangle$ and back by a second $\pi$-pulse on the same transition. The accumulated $2\pi$-pulse yields the initial state with a minus sign if the atom actually occupies the state $|1-b_i\rangle$ but no phase change if the atom occupies the opposite state, and this operation thus provides the conditional phase  component of the Grover search on a one-bit register.

We are interested in testing whether the input state simultaneously fulfills agreement with all bits in the comparison and here the Rydberg blockade is useful. Fig.1.c) summarizes the basic idea: if the control atom $i$ is excited to a Rydberg state, the excitation energy of the Rydberg state of the target atom $j$ is raised by the strong dipole-dipole interaction between the atoms, and hence excitation by a laser which is resonant with the single target atom transition is blocked. A $2\pi$ pulse on the target atom thus has two possible outcomes: no effect if the control atom occupies the Rydberg state and a change of sign if the control atom does not occupy the Rydberg state.

Applying the $\pi$-pulse $|1-b_0\rangle \rightarrow |r_0\rangle$  will result in excitation of the components of the register quantum superposition state where the $0^{th}$ qubit does not match the marked element. One proceeds with a $\pi$-pulse $|1-b_1\rangle \rightarrow |r_1\rangle$ on the $1^{st}$ atom. Note that due to the blockade, this pulse has no effect on state vector components where the $0^{th}$  atom is already excited, but components which did match the marked element on the $0^{th}$ bit, and which were therefore not excited, will now become excited if they do not match the $1^{st}$ bit. This is illustrated in Fig. 2, where we use the symbol $R_{0,r}(\pi)$ to denote a $\pi$-pulse excitation between states $|0\rangle$ and $|r\rangle$, and similarly for state $|1\rangle$. The figure shows how we attempt to successively excite each atom from the ideally unpopulated state $|1-b_i\rangle$ to the Rydberg state, and eventually all input state components have undergone a unitary transform into new states with precisely one Rydberg excited atom (the first atom in the register not fulfilling the comparison with the desired bit string) or no Rydberg excitation at all in case all atoms occupy the appropriate ground state $|b_i\rangle$ defined by the marked element. Applying now a second set of $\pi$-pulses in reverse order, starting with the  $k-1^{st}$ atom and finishing with the $0^{th}$ atom, we return all atoms to their initial state, and we equip all components that actually underwent a $2\pi$ transition to the Rydberg state and back with a change of sign. Note that if we had intertwined a resonant Raman process on an ancilla target atom between states $|0\rangle$ and $|1\rangle$ via the Rydberg state between the two sequences of $\pi$-pulses, the net effect of the operation would have been a $k$-atom controlled NOT operation on that atom, controlled by the agreement of the $k$-bit atomic register with the bit string $b_0, ... b_{k-1}$, originally proposed by Isenhower and Saffman \cite{Isenhower2010}.

\begin{figure}
  \includegraphics[width=8cm]{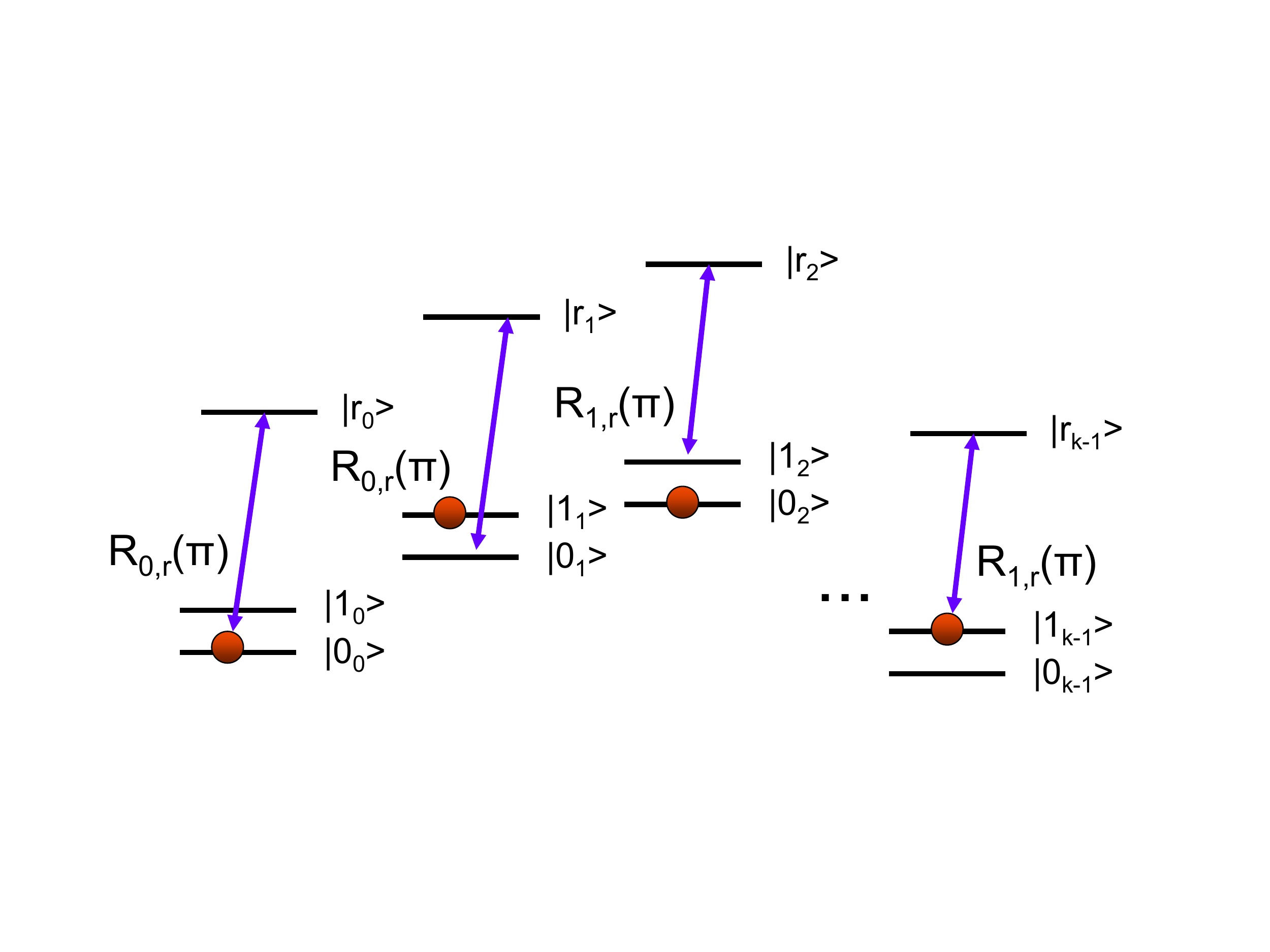}\\
  \caption{Grover conditional phase: Successive $\pi$-pulse excitation transfer of all atoms from one of their qubit states to the Rydberg state. The lasers couple the states $|1-b_i\rangle$, complementary to the bit values of the marked element $x_0$, to the Rydberg states. If one atom is already excited, no further excitation occurs, and hence quantum register components with more atoms populating the states coupled to the laser field will only be excited once, while the state with no atoms coupled to the laser fields will not be excited at all. A second set of $\pi$ pulses, applied to the atoms in reverse order, returns all population to the initial states, but a relative change of sign has been accumulated between the state $|x_0\rangle$ and all other components. }\label{Fig.2}
\end{figure}

Since the Rydberg blockade ensures that components of the quantum state with any number of register values differing from the $b_i$'s will be excited and de-excited at precisely one location (atom) in the register, while the $|x_0\rangle$ component is left unchanged, the whole operation can be written
\begin{equation}
|x\rangle \rightarrow \{
\begin{array}{c}
  |x\rangle, (x=x_0) \\
  -|x\rangle, (x\neq x_0).
\end{array}
\end{equation}
Apart from an irrelevant global change of sign, this accomplishes the conditional sign step of the Grover algorithm.

\subsection{Inversion about the mean with Rydberg blockade interaction}

The inversion about the mean (\ref{GroverInv}) is given by a simple expression in the $x$-representation.  If we stay in the Hilbert space of dimension $N=2^k$, the transformation  (\ref{GroverInv}) is written, $|\Psi\rangle \rightarrow U_G|\Psi\rangle$, with the matrix
\begin{equation}
U_G = 2 P - I,
\end{equation}
where $P$ is the $N\times N$ matrix with the value $1/N$ in all positions and $I$ is the $N\times N$ identity matrix. Defining the initial state of the search algorithm with uniform amplitudes, $|\Psi_0\rangle \equiv \sum_x \frac{1}{\sqrt{N}} |x\rangle$, we observe that $P$ is the projection operator,
\begin{equation}
P = |\Psi_0\rangle\langle \Psi_0|.
\end{equation}
Note that we can write
\begin{equation}
U_G=P-(I-P)=P-Q,
\end{equation}
where $Q$ is the projection on the space orthogonal to $|\Psi_0\rangle$. Writing the operation like this, we recognize that this is also, apart from a global phase factor, a conditional change of sign, conditioned on the state not being $|\Psi_0\rangle$.

The quantum algorithm is applied on an atomic quantum register, where different input values are encoded in binary representation in the individual ground state atomic manifolds. The uniform input state $|\Psi_0\rangle$ and the associated projection operator
$P$ is so far expressed in the basis of the variables $x$, but it is  also straightforward to identify the expansion of the state in binary representation, where $|\Psi_0\rangle$ is simply a product state of all qubits, prepared in the superposition state $(|0\rangle+|1\rangle)/\sqrt{2}$. To multiply a phase factor on the state component along $|\Psi_0\rangle$ is thus equivalent to performing this multiplication conditioned on all qubits being in the state $(|0\rangle+|1\rangle)/\sqrt{2}$. This is, indeed, equivalent to our task in the previous section, except that in that section, we checked the occupation of the states $|1-b_i\rangle$ complementary to the desired classical bit value $b_i$, while now, we must check for the occupation of a superposition state.

Fig.3 illustrates how in the second Grover step, excitation from qubit superposition states is driven by two fields in a $\Lambda$-configuration with well defined phases and amplitudes, while in the first Grover step, shown in Fig.2, checking for the bit values $b_i$ uses only a single field from the $|1-b_i\rangle$ state. It is interesting to note, that the $\pi$-pulses applied always occur (or do not occur) from the ground state manifold to the Rydberg state and back, and not between superpositions of these states. This implies that a variety of pulse sequence schemes and chirped adiabatic passage techniques can be applied to make these pulses robust against small disturbances of the system and variations in the field parameters, and this robustness also applies to qubit superposition states in the ground state manifold \cite{Roos2004}.

\begin{figure}
  \includegraphics[width=8cm]{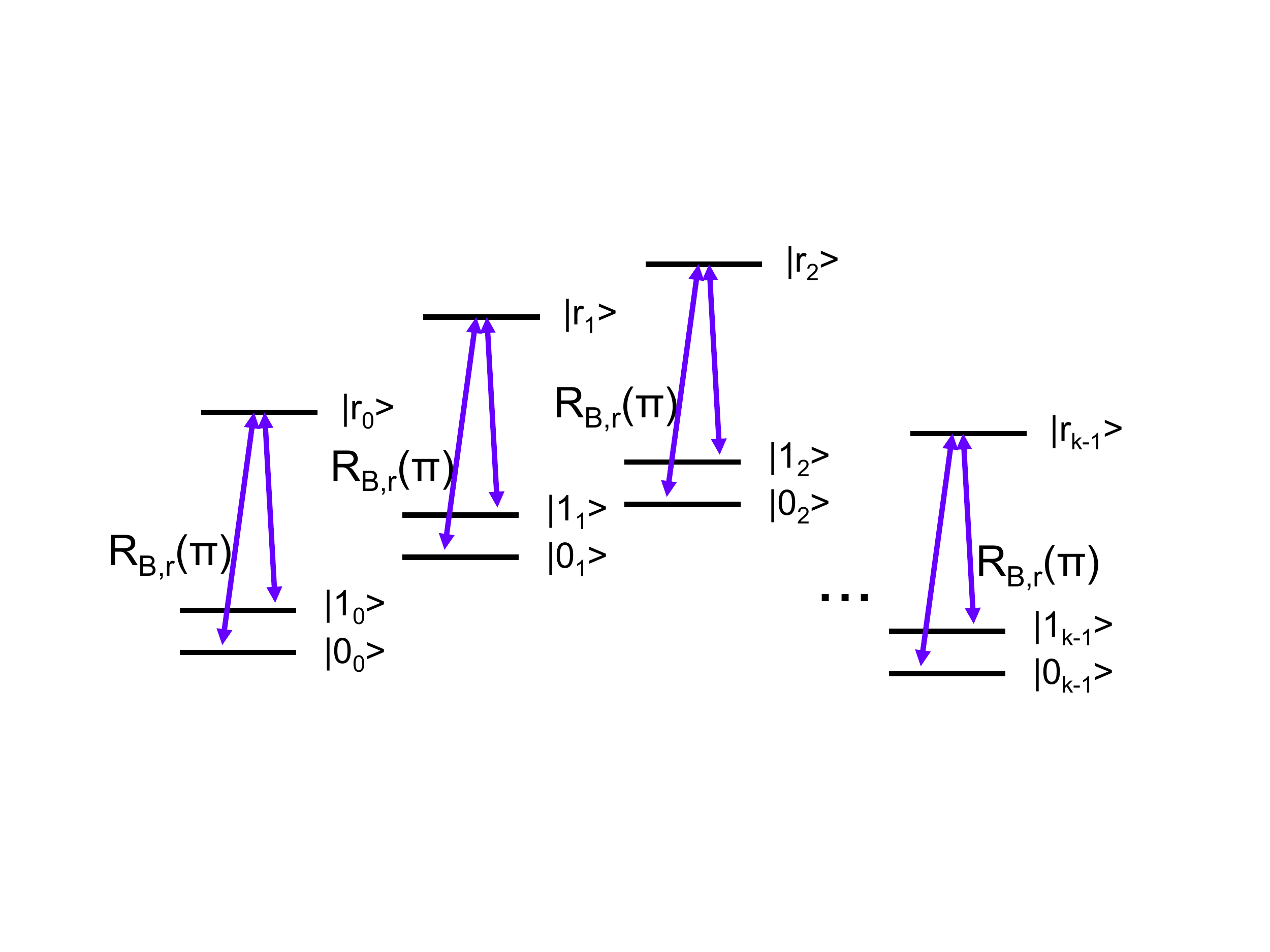}\\
  \caption{Grover inversion-about-the-mean: With the $\Lambda$-transition laser excitation scheme shown, the $(|0_i\rangle+|1_i\rangle)/\sqrt{2}$ dark state in each atom is uncoupled while a $\pi$-pulse excitation transfers the bright superposition qubit state $|B_i\rangle=(|0_i\rangle-|1_i\rangle)/\sqrt{2}$ to the Rydberg state. In a succession of such excitation pulses, if one atom is already excited, no further excitation occurs, and hence quantum register components with more atoms populating the bright states will only be excited once, while the state $|\Psi_0\rangle$ with no bright state atoms will not be excited at all. A second set of $\pi$ pulses, applied to the atoms in reverse order, returns all population to the initial states, and a relative change of sign has been accumulated between the state $|\Psi_0\rangle$ and all other components. }\label{Fig.3}
\end{figure}

The projection of the full register state on $|\Psi_0\rangle$ vanishes if just one atom $i$ does not occupy the state $(|0_i\rangle+|1_i\rangle)/\sqrt{2}$, and we can hence check the state bit by bit by exciting the Rydberg state with two laser fields in a $\Lambda$-configuration, for which $(|0_i\rangle+|1_i\rangle)/\sqrt{2}$ is a dark state, i.e., with two Rabi frequencies of same magnitude and opposite sign \cite{Roos2004}. This laser configuration excites the bright state $|B_i\rangle=(|0_i\rangle-|1_i\rangle)/\sqrt{2}$ into the Rydberg state and serves the same purpose as our excitation of the $|1-b_i\rangle$ register state in the first step of the algorithm. As above, we thus begin the sequence of operations by exciting the $0^{th}$ atom coherently from the bright state to the Rydberg excited state $|r_0\rangle$. Thereafter, we excite the $1^{st}$, and subsequently all the following  atoms on the same transition, noting that for any component in the full register superposition state the first atom  in the bright state will be excited and will hereafter block excitation of any further atoms in the same state component. The net result of the sequence of excitation steps is that the $|\Psi_0\rangle$ component of the register is left unchanged, while all other components have precisely one Rydberg excited atom. Applying the same $\pi$-pulses, but now in opposite order to the atoms in the register, will return all Rydberg excited state population back to the ground state where it came from, but writing also a minus sign on the quantum state. So, after the pulse sequences, all population is restored into the ground state qubit space, but all components of the state vector except $|\Psi_0\rangle$ have undergone a change of sign. This is precisely the unitary operation needed for the Grover inversion-about-the-mean!

It is worth pointing out, that by leaving the ground state qubit space and carrying out multiple operations on several atoms before finally returning to the qubit space again, the entire multi-bit gates described in this and the previous subsection are explicitly not formed as sequences of one- and two-qubit gates. The Rydberg blockade interaction between all qubits provides a shortcut to multi-bit operations, which may not be found in other quantum computing implementations, where other shortcuts may apply instead, \textit{e.g.},  via the centre-of-mass motional degree of freedom in quantum computing with a string of trapped ions \cite{Blatt2008}. Indeed, some of the ingredients in the above analysis of the Grover algorithm were already discussed and can be implemented with quite different interaction steps in the ion trap computer \cite{Wang2001} and in cavity QED based proposals where a single field mode couples to all qubits \cite{Scully2001}.

\section{Alternative configurations and interaction schemes}

Favorable conditions for Rydberg blockade are associated with the F\"orster resonance phenomenon occurring when there is a near degeneracy between the product state with two Rydberg excited atoms and another Rydberg product state \cite{Protsenko2002,Saffman2005,Walker2008}. None of the single electron states have mean dipole moments, but the non-vanishing dipole matrix elements between the different single atom states couple the two product states and split their energy levels. The resonant F\"orster resonance mechanism gives rise to large energy shifts scaling as $1/R^3$ with the interatomic distance, approaching $1/R^6$ for very large distances, where the coupling gets smaller than the energy difference between the two states. Being relatively isotropic for $s$ states, the F\"orster mechanism is useful to provide coupling between any pair of atoms in a not too large system of particles.

\subsection{Simultaneous excitation of all atoms}
\label{sec.simultaneous}

It is tempting to look for a protocol which uses the same excitation mechanisms of the individual atoms, but applies the coupling lasers at the same time to all atoms. In both steps of the Grover algorithm, there is a state component which is not excited at all, and this component would also be invariant under the simultaneous application of the laser fields to all atoms in the register. The state components which differ from the register state $|x_0\rangle=|b_0 b_1 ... b_{k-1}\rangle$ (or from the uniform state $|\Psi_0\rangle)$ at exactly one qubit location, will be excited precisely at that single location under the simultaneous interaction with the laser fields, and after the subsequent deexcitation, the register acquires the desired phase factor for these components. If driven simultaneously, the state with two ``erroneous" qubits, $i,j$, experiences a coupling towards the state with two Rydberg excited atoms.  That state, however, is shifted by the F\"orster interaction, and what happens, instead, is that the ground state, say $|g_i,g_j\rangle$ couples to a symmetric combination $(|r_i,g_j\rangle + |g_i,r_j\rangle)/\sqrt{2}$ with only one Rydberg excited atom. The Rabi frequencies of the two coupling terms interfere and lead to  an effective coupling which is $\sqrt{2}$ larger than the single atom excitation Rabi frequency \cite{Gaetan2009}. If three atoms occupy states, simultaneously coupled to the Rydberg states, the enhancement is $\sqrt{3}$, etc, and since the superposition state of the register contains components with all these different numbers, it is not possible to apply a simple constant laser pulse, which will act as a $\pi$ pulse on all the transitions.
One may imagine a pulse sequence, identified by optimal control theory, which by suitably varying the amplitudes and phases of the fields, yields an effective $\pi$-pulse on all or many of the atomic components with different Rabi frequencies. When designing such a pulse sequence one must ensure that it does not widen the resonance band width beyond the Rydberg blockade interaction shift and hence permits doubly occupied Rydberg states. Sequential addressing of the individual atoms may ultimately be both faster and more robust than such a collective stategy.

A more realistic scheme including simultaneous excitation of all register atoms can be engineered if excitation is possible to two different Rydberg states, $|s\rangle$ and $|r\rangle$, with the property that a pair of state $|s\rangle$ atoms is not near degeneracy with any other pair of states (no F\"orster resonance), while a pair occupying $|r\rangle$ and $|s\rangle$ does constitute a dipole-dipole resonant state and hence experiences a large blockade shift. Possible states in rubidium with this property are $|s\rangle=|40 p_{3/2},m=1/2\rangle$ and $|r\rangle=|41 s_{1/2},m=1/2\rangle$ \cite{Saffman2009}.

Some calculations indicate that the dipole interaction shifts and hence the blockade mechanism are generally present also for more than two excited atoms \cite{BrionMM2007}, but special cases have been identified where two atoms block, while among three atoms resonant excitation of more than a single atom may be allowed \cite{Pohl2009}. We will assume that one can avoid the latter phenomenon by appropriate choice of atomic systems, but we note that it may take a more detailed calculation to precisely assess the collective Rydberg interaction shifts within a collection of state $|s\rangle$ excited atoms and a single $|r\rangle$ atom.

Now, in this situation one can simultaneously drive $\pi$-pulse excitation pulses on all atoms from the relevant single qubit states to the Rydberg state $|s\rangle$, which thus gets a total occupancy of zero for the state component $|x_0\rangle$ and $|\Psi_0\rangle$ in the first and second part of the Grover step, respectively, and a total occupancy of unity or more (!) for all other components.
We assume that a single ancillary atom is present in the state $|0\rangle$ and has not yet been excited by laser fields. This atom is now driven towards the Rydberg state $|r\rangle$, and due to the F\"orster resonance interaction, this transition is blocked if there is already any number  of atoms $(\geq 1)$ present in state $|s\rangle$, see Fig.4.

\begin{figure}
  \includegraphics[width=8cm]{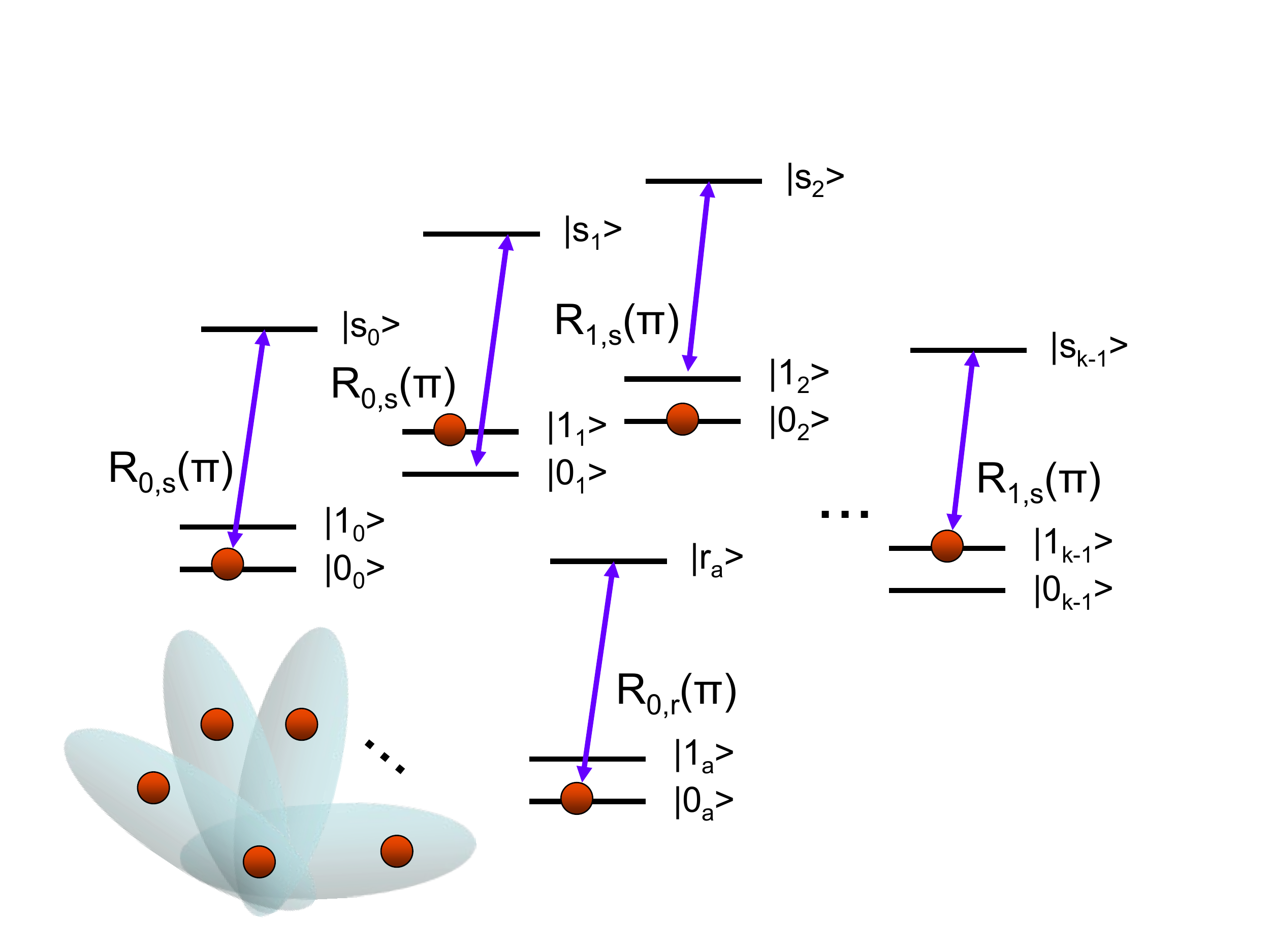}\\
  \caption{Grover conditional phase: Simultaneous $\pi$-pulse excitation transfer of all atoms from a specific qubit state to the Rydberg state $|s\rangle$. The $|s\rangle$ state atoms do not interact with each other, and the number of Rydberg excited atoms is not limited to zero and unity. Finally one ancilla atom, prepared in state $|0\rangle$ is excited by a $2\pi$-pulse towards the Rydberg state $|r\rangle$, which does interact with all $|s\rangle$-state atoms (see the insert sketch of the atomic mutual interactions). A second set of $\pi$ pulses, applied to all atoms in reverse order, returns all atoms to their initial states, but a relative change of sign has been accumulated between the state with no Rydberg excitations and all other components. Exciting the qubits from the states complementary to the bit values $b_i$ thus yields the conditional phase operation, while excitation from the bright qubit superposition states $|B_i\rangle=(|0_i\rangle-|1_i\rangle)/\sqrt{2}$ yields the inversion about the mean.}\label{Fig.4}
\end{figure}

Completing a full $2\pi$ pulse on the $|0\rangle - |r\rangle$ transition of the ancillary atom thus yields a change of sign if the register initially occupies the targeted state $|x_0\rangle$ (first Grover step) or $|\Psi_0\rangle$ (second Grover step), and no process and no change of phase occurs otherwise. Transferring finally all atoms in state $|s\rangle$ simultaneously back to their initial ground state by inversion of the first $\pi$-pulses thus completes the operation (note that we invert the $\pi$-pulses by laser fields with the opposite sign of the first exciting pulses - in this way, we do not accumulate sign changes of the state vector). A single iteration of the conditional phase shift and the inversion about the mean is thus accomplished in the time it takes to perform four single atom $\pi$-pulses on the ground state to $|s\rangle$ transitions and two single atom $2\pi$-pulses on the ground state to $|r\rangle$ transition. In a database with $N$ elements, one has to repeat these operations $\sim \sqrt{N}$ times to complete the full Grover search algorithm.

\subsection{Grover search with a sub-register architecture}
\label{sec.subreg}

To scale the system to large numbers of qubits, it is difficult to retain individual addressability while still finding room for all the atoms within the Rydberg interaction range of each other or of a central ancilla particle. In this section we will show, that it is then possible to split the system in separate sub-registers, which each constitute a cluster around an ancilla atom, as  sketched in Fig.4. As in Fig.4, we assume no F\"orster resonance and hence no blockade effect associated with atoms in Rydberg state $|s\rangle$, while any non-zero number of logical qubit atoms excited to $|s\rangle$ will block the excitation of the ancilla qubit to the Rydberg state $|r\rangle$. At each query and inversion step, we start with all the ancilla atoms in state $|0\rangle$, we excite the sub-register atoms to the state $|s\rangle$ as in the previous section, and we transfer the ancilla atoms via $|r\rangle$ to the state $|1\rangle$. The relevant property of the sub-ensemble logical qubits is thus encoded in the population of this final ancilla state rather than in a phase shift. The logical qubit atoms are then transferred back to their respective ground states by inverting the $\pi$-pulse excitation processes applied to each atom, and the task is now to compare the ancilla atoms.

Changing the sign of the quantum state with all ancillas occupying state $|1\rangle$ while retaining the sign on all other states would be easy if all ancillas were within the Rydberg blockade radius of each other: it would merely consist in applying the query operation, described in Sec. II.A with the \textit{ancilla register } marked element $x^a_0=11 ... 1$ . This query of the ancillas could be
obtained by physically moving the atoms from their sub-register locations to a tighter interaction volume, in the same spirit as the ion shuttling architecture \cite{WinelandShuttle}. Let us present, however, a scheme that will work with stationary ancilla atoms which  experience only a finite Rydberg interaction strength $\Delta E_{rr}$, which is too weak to exercise a reliable excitation blockade.  With this interaction, the internal state of one atom can control a transfer process in another atom in the following way: \textit{i)} the control atom is excited and the target atom is excited from a (bright) superposition $(|a\rangle-|b\rangle)/\sqrt{2}$ of the initial and desired final states $|a\rangle$ and $|b\rangle$ to the Rydberg state $|r\rangle$, \textit{ii)} a change of sign accumulates on the $|rr\rangle$ product state component and \textit{iii)}  inverse pulses return the Rydberg state amplitude to the ground states.  The resulting controlled sign change on the bright superposition state of the target atom is equivalent to a transfer between the initial and final state \cite{Roos2004}, for example,
\begin{eqnarray} \label{int-gate}
|a\rangle = \frac{1}{\sqrt{2}} \big((|a\rangle+|b\rangle)/\sqrt{2}+(|a\rangle-|b\rangle)/\sqrt{2}\big)\nonumber \\
\rightarrow \frac{1}{\sqrt{2}} \big((|a\rangle+|b\rangle)/\sqrt{2}-(|a\rangle-|b\rangle)/\sqrt{2}\big)=|b\rangle.
\end{eqnarray}
While not each ancilla atom needs to interact with all other ancilla atoms, we assume that we can pick a sequence of operations involving always pairs with sufficient Rydberg interaction to accumulate a change of sign, and hence the controlled state transfer, after a dwell time $\tau = \pi/\Delta E_{rr}$ in the Rydberg state of a few microseconds.

We propose to verify the ancilla quantum states in a binary search tree pattern, so that we group the ancillas in pairs, encode their joint state into the first atom of each pair, form new pairs of these ancilla atoms for which the joint state is again encoded in the first atom, and repeat until one final atom occupies state $|1\rangle$ if and only if all the ancillas were put in state $|1\rangle$ by their sub-register qubits. A  $2\pi$ pulse on the last ancilla atom provides the desired conditional phase, and reversing all ancilla operations through all branches of the search tree, we return all ancillas to the initial state $|0\rangle$, retaining the conditional phase factor only on the marked element in the query operation and on the corresponding component in the inversion about the mean operation.

Two ancilla atoms may populate four logical states, $|00\rangle,\ |01\rangle, |10\rangle$, and $|11\rangle$, for which we want the first atom to end up in state $|0\rangle$, $|0\rangle$, $|0\rangle$, and $|1\rangle$, respectively. Since we want to apply reversible and unitary operations, for three product states to end up with the same state of the first ancilla atom, the second ancilla must accomodate three orthogonal states. We thus assume ancilla atoms with a three level structure $|0\rangle,\ |1\rangle,\ |2\rangle$, and note that an operation which transfers $|10\rangle$ to $|02\rangle$ and retains all other product states will be adequate. That operation can be done in two steps, $|10\rangle \rightarrow |12\rangle \rightarrow |02\rangle$, which are both of the controlled transfer type (\ref{int-gate}): \textit{i)} the state $|1\rangle$ of the first atom controls the transfer between $|0\rangle$ and $|2\rangle$ of the second atom, \textit{ii)} the state $|2\rangle$ of the second atom controls the transfer between $|1\rangle$ and $|0\rangle$ of the first atom. By going through all four possible input states, one verifies that the operations indeed perform the desired task.

As we set out to achieve, the first ancilla atom in each pair now encodes the logical AND of the two atoms, and by forming the new pairs and repeating the operations a total of  $n_s/2 + n_s/4 + n_s/8 ... = n_s-1$ times, a single atom witnesses the state of all $n_s$ sub-ensembles. We invert all steps by the same operations in reverse order, and hence a single query or inversion about the mean on the full register, involves 2 blockade gates between the qubit atoms and the ancilla atom within each sub-register ($2n_s$ operations carried out sequentially or in parallel), and $2(n_s-1)$ interaction gates between pairs of ancilla atoms (which can also be done in fewer simultaneous operations on different pairs).

\section{Scaling of errors}

The search size that can be implemented using multibit gates depends on the error scaling with $k.$ A complete analysis of this question, even in the ideal situation of perfect hardware, would require a simulation of the multibit gate that accounts for errors due to spontaneous emission from Rydberg levels, finite blockade efficiency, and the multiplicity of high $n$ Rydberg levels. Such an analysis is cumbersome and  has not yet been performed, even for the much simpler case of a two qubit CNOT gate.
We will instead follow the approach used in \cite{Saffman2005,RMP2010} and assume the
gate error averaged over all possible input states is small, so that the error can be well approximated by separately adding errors due to Rydberg state spontaneous  emission and finite blockade strength.
When such an analysis is performed for a controlled phase gate it is found that
the error is minimized by choosing a Rabi frequency $\Omega_{\rm opt} = (7\pi)^{1/3} {\sf B}^{2/3}/\tau^{1/3}$ which gives a minimum gate error
 $E = (3(7\pi)^{2/3}/8) ({\sf B}\tau)^{-2/3}.$ Here $\sf B$ is the two-atom blockade interaction strength and $\tau$ is the Rydberg spontaneous lifetime.

Applying the same type of analysis to the Grover phase inversion step leads to equivalent scaling  $\Omega_{\rm opt}\sim {\sf B}^{2/3}/\tau^{1/3}$ and $E\sim c(k) ({\sf B}\tau)^{-2/3}$ with a numerical prefactor $c(k)$ which is approximately linear in the number of
bits $k$. A detailed accounting of the analysis leading to this result will be presented elsewhere \cite{2011}. In the present context it is interesting to note that
there are qualitative differences in the
error scaling of the multibit gates as compared to that of the two-bit CNOT gate, despite the mathematical similarity of the result.  For the two-bit blockade gate intrinsic errors $<10^{-4}$ are theoretically possible by exciting Rydberg levels with $n>100$ \cite{Saffman2011} at small atomic separations of $R<5~\mu\rm m$. Low errors can be obtained at larger separations by going to higher $n$. At large $R$ we have a van der Waals interaction and  ${\sf B}\sim n^{11}/R^6, \tau\sim n^3$ so $E\sim n^{-28/3} R^4.$

In contrast with this situation, implementation of a multibit interaction leads to some additional constraints. Let us suppose that $k$
atoms are arranged on a two-dimensional square lattice with $\sqrt{k}$ atoms on a side and period $d$. Ideally we want to have a strong interaction for atoms with the maximal spacing of $R_{\rm max}=\sqrt{2(k-1)} d$. We may choose $n$ large enough to ensure an adequate ${\sf B}(R_{\rm max}).$  However, doing so will result in
${\sf B}(d)$ being a factor $(2(k-1))^3$ larger for van der Waals scaling.
  If the resulting ${\sf B}(d)$ exceeds half of the spacing between Rydberg levels of neighboring $n$ then the blockade can actually be reduced for proximal atoms, which will lead to large errors. To avoid this from happening we must respect  ${\sf B}(d)<[U(n)-U(n-1)]/2\sim n^{-3}.$ Thus the error scaling for a multibit gate is
$E\sim (n^{-3}\times n^3)^{-2/3}\sim ~\rm constant$. The implication is that the principal quantum number $n$ should be chosen large enough to ensure that the entire array is in the resonant dipole-dipole limit $({\sf B}\sim 1/R^3)$ to minimize interaction strength variations across the array, but once we have met that constraint there is no advantage to going to higher $n$.

We have analyzed the fidelity of a Grover phase inversion step, including lattice averaging of the different powers of the interaction strength which arise at intermediate stages in the error analysis.
It turns out that it is not possible to limit the maximal interaction at spacing $d$ and simultaneously have the entire lattice in the ${\sf B}\sim 1/R^3$ regime for $k\ge 9$. We have therefore averaged the actual computed values of $\sf B$ over the lattice.
The errors for  different architectures are given in Table \ref{tab.errors}.
 For Cs $|ns\rangle$ states  with $n=75$ we find that for search of a 512 element database  $(k=9)$ the average
error is $E\simeq 0.002.$ The error for inversion about the mean will be
essentially the same, so one Grover iteration has an expected error of $E\simeq 0.004$. The error for 65536 elements ($k=16$) is  $E\simeq 0.015$.  It should be emphasized that these are average errors and  the actual error for some input states may be much larger.

The error scaling is different for the simultaneous excitation approach presented in Sec. \ref{sec.simultaneous}. In this case we require a large asymmetry between the
$|s\rangle - |s\rangle$ (weak) and $|s\rangle - |r\rangle$ (strong) interactions.
We choose states and lattice spacings such that $|s\rangle - |s\rangle$ is in the van der Waals regime and scales as $n^{11}$ while $|s\rangle - |r\rangle$ is a resonant dipole-dipole interaction scaling as $n^4$. The ratio is thus maximized by keeping $n$ not too large. A possible choice with Cs atoms is $|s\rangle=|60p_{3/2}\rangle$ and $|r\rangle=|60s_{1/2}\rangle.$ Using a square lattice with the ancilla atom at the center we will have $k-1$ atoms available for register encoding. Averaging the $|s\rangle - |r\rangle$ interaction over the lattice\cite{2011} we find
the error for each full Grover step (controlled sign plus inversion about the mean) is (0.04,0.14,0.20) for $k-1=(8,15,24).$ Although these intrinsic errors are a
few times larger than for the individual addressing protocol, this may be partially compensated by the technical advantage that  global addressing requires $2/k$ fewer Rydberg pulses.

\begin{table*}
\centering
\begin{tabular}{|l|c|c|c|c|c|c|c|}
\hline
& & &sequential& simultaneous&simultaneous with&quadratic speedup \\
 $N$& $k-1$&$k$& addressing & addressing & sub-registers&limit $N^{-1/4}$\\
\hline
256 & 8 &&&.08&&.25\\
512 & &9 &.004&&&.21\\
32768 & 15 &&&.20&&.074\\
65536 && 16 &.015&&$ .16~(k_s=8, n_s=2)$&.063\\
16777216 & 24 &&&.28&$ .24~(k_s=8, n_s=3)$&.016\\
\hline
\end{tabular}
\caption{
\label{tab.errors}Errors per Grover iteration step for the different architectural approaches described in the text. }
\end{table*}

The error scaling for the  sub-register architecture presented in Sec.
\ref{sec.subreg} can be estimated as follows. Denoting the error in the simultaneous excitation approach with $k$ bits as $E(k)$ the
sub-register architecture error for each Grover iteration is
$E_{\rm sr}=n_s E(k_s) + (n_s-1) E_{\rm a}$ where $k=n_s k_s$ and $E_{\rm a}$ is the additional error associated with each of the ancilla comparison operations. These ancilla comparison operations rely on interaction gates, not the blockade gates we use elsewhere. Inspection of Fig. 15 in \cite{RMP2010} shows that using, e.g. Rubidium 150s states will allow these errors to be  less than $0.01$ at distances out to $100~\mu\rm m$. The physical separation of two centrally positioned ancilla atoms in neighboring sub-registers will be roughly  $d\sqrt{k_s} $ so for $d<10~\mu\rm m$ and $k_s<25$ we are working with a distance that is under $50~\mu\rm m$.  Referring to Table \ref{tab.errors} we see that the error overhead associated with the ancilla comparison steps is therefore negligible for $k>8.$ Since the simultaneous addressing error grows faster than linearly with $k$ the sub-register architecture can be advantageous and leads to the potential for implementation on  rather large search problems.

  Although the errors shown in the Table are too large  for standard circuit model fault tolerant operation they may be adequate for Grover search, where the idea is to increase the amplitude of a desired component of the wavefunction. An error in one step may not lead to the desired amplification, or may even push the state vector in the wrong direction. Nevertheless, as long as the average error is sufficiently small we expect the search procedure to proceed towards  the sought after solution. It has been shown in \cite{Shenvi2003} that a quadratic speedup is still possible provided the oracle phase
error per step is bounded by $N^{-1/4}$. For larger errors the quantum speedup  will be less than quadratic.  Comparison of the listed errors with the last column in Table \ref{tab.errors} shows that
a full quadratic speedup is possible for problems as large as $N=65536$ with sequential addressing.

We note that the sequential addressing scheme generally has lower  errors than  simultaneous addressing, with or without sub-registers. This conclusion is based on ignoring technical errors associated with imperfect laser pulses. Since the simultaneous schemes require a factor of $k/2$ fewer laser pulses they may nevertheless be advantageous. A reliable comparison of the approaches, as well as an accurate prediction of the possible speedup, will depend on implementation details and is beyond the scope of this work.

\section{Conclusion}

In this paper we have suggested to use the particular multi-atom possibilities of the Rydberg blockade interaction to implement the Grover search algorithm using  multi-bit quantum gate operations rather than the conventional one- and two-qubit circuit model. The Rydberg blockade mechanism ``takes the atoms to a third level" (the Rydberg state) and hence leaves the qubit and quantum circuit paradigm for an extended number of operations during gate operation. The construction of multi-bit operations and entire algorithms  without the use of predefined universal gates is not viable or meaningful as a general approach to quantum computing as it merely represents a transfer of the computational complexity, which should ideally be taken care of by the quantum processor, to the theoretical design of the gates. In the present case, however, the experimental strategy is so straightforward that simple schemes are readily derived.

We focussed on the Grover algorithm, showing first that it can be decomposed into sign changes conditioned on all register qubit states, and secondly that these operations can be accomplished by $\pi$ pulse excitations of the individual atoms. A general $k$-qubit unitary operation is described by a $2^k\times 2^k$ matrix, and the theoretical lower bound on the number of $C-NOT$ gates needed, together with arbitrary single-qubit gates, to form an arbitary unitary operation scales as $4^k$ \cite{Shende}. A practical, constructive protocol reaching this limit is presented, e.g., in \cite{Mikko2004}. In comparison, a quantum circuit design has been made which decomposes the Grover step into $49k-149$ elementary one- and two-bit gates ($k>3$) \cite{Zubairy2002}. Our algorithm requires $\sqrt{N}$ repetitions of the Grover step (controlled sign and an inversion about-the-mean), and we showed that in physical situations where the atoms must be excited sequentially this step requires $4k$ single atom $\pi$ pulses, while for level schemes offering both interacting and non-interacting atoms in the Rydberg states, they can be accomplished in the duration of only  $8$ single atom $\pi$-pulses. Furthermore, the estimated fidelity benefits from the multi-bit gate operations and is suggestive that the Grover algorithm  can be successfully implemented on a quite large register.

We are convinced that our ideas for multi-bit gates, which go just a little bit beyond the qubit and circuit-model paradigm may be pursued much further, and although they may not change the speed-up offered by quantum computers in a conceptual way, they may offer practical and also quite efficient ways to carry out quantum computing protocols in practice.

This research was supported  by the EU integrated project AQUTE, the IARPA MQCO program, and DARPA and NSF award PHY-0969883.

\end{document}